\documentclass[aps,prl,twocolumn,showpacs,superscriptaddress,groupedaddress]{revtex4}  % for review and submission
\usepackage{graphicx}  % needed for figures
\usepackage{dcolumn}   % needed for some tables
\usepackage{bm}        % for math
\usepackage{amssymb}   % for math
\usepackage{color}

\newcommand{\sech}{\mathop{\mathrm{sech}}}

\newcommand{\diag}{\mathop{\mathrm{diag}}}
\newcommand{\sgn}{\mathop{\mathrm{sgn}}}

\newcommand{\ppsi}[2]{{\psi_{#1}^{(#2)}}}

\usepackage{amsmath}
\usepackage{mathtools}
\usepackage{tabularx}
\usepackage{fancyhdr}
\usepackage{datetime}
\usepackage{bbold}
\usepackage{enumitem}
 \newcommand{\tendsto}[1]{%
  \xrightarrow{\smash{\raisebox{-2ex}{$\scriptstyle#1$}}}}

\def\abs#1{\left\lvert#1\right\rvert}

\begin{document}
\title{Radiation pressure in the Coupled Nonlinear Schr\"odinger Equation exerted on the solitons.}
\author{Tomasz Roma\'nczukiewicz}
\affiliation{Institute of Physics, Jagiellonian University, Krak\'ow, Poland}

\begin{abstract}
In the paper an interaction between a soliton and a traveling wave in the Coupled Nonlinear Schr\"odinger Equation (CNLSE) is considered. 
The force acting on the soliton can both push or pull the soliton depending on parameters of the equation, structure of the soliton and
the incident wave. In special cases analytical results are presented and in general numerical solutions of both linearized and the full problem are 
shown.
\end{abstract}

\pacs{03.75.Lm, 05.45.Yv,42.65.Tg}
\maketitle

\textit{~~Introduction.~~}
The coupled nonlinear Schr\"odinger equation (CNLSE) appears in many physical problems starting from propagation of wave packets in wave guides in 
nonlinear optics \cite{YangTan, Goodman, Kivshar, Kaup} to description of the Bose-Einstein condensates \cite{Malomed, Sacha1, Sacha2, Sacha3}.
One of the most interesting solutions of the equations are solitons, studied extensively mostly in the NLSE due to its integrability. 
The coupling breaks the integrability opening new possibilities such as for example resonant multi bounce collisions \cite{YangTan, Goodman}.\\
In our previous papers we have found that solitons in various models reveal a surprising feature which we call negative radiation pressure 
(NRP). 
When small waves (radiation) hit a large object, such as a soliton, they scatter and exchange momentum and energy. 
Usually this leads to a force pushing  the object. 
However there are situations when the force exerted by radiation pulls the object. 
This feature is present in many well known and extensively studied models including kinks in $\phi^4$ model \cite{mypapers1,Forgacs}, kinks (and 
domain walls) in three vacuum $\phi^6$ model \cite{Patrick, phi6}, vortices in abelian, nonabelian Higgs models \cite{vortices} and in the 
Gross-Pitaevskii (GP) model \cite{tractorBeams}. 
This feature requires some mechanism which transforms energy from slower modes to faster ones.
In the $\phi^4$ model nonlinear coupling creates a double frequency wave which effectively exerts NRP on the kink. This is possible due to a 
very rare feature of reflectionless  of the $\phi^4$ kink in the first order. 
However we have found that there exists a second mechanism which can reveal in many important physically models. 
We have shown that sometimes it is enough that there exist two modes with different masses or, in more general, different dispersion relations.
As an example asymmetric kinks and domain walls in $\phi^6$ model can serve, which we discuss in a separate paper \cite{phi6}.
Waves coming from one side exert negative radiation pressure and from the other side positive.
Similar mechanism can be observed in a theory of more than one component field. Global $U(1)$ theory exhibits a vortex which is a topological defect 
with a winding number as a topological charge \cite{vortices}. 
Moreover in such theory there exist two types of small excitation of a vacuum: a massless Goldstone's mode and a massive mode responsible for excitation of the amplitude of the 
field. 
A scattering on a vortex can lead to a mixture of the two modes. The mass difference can lead to NRP in case when the amplitude wave hits the 
vortex.\\
Similar phenomena are recently described in other fields and sometimes referred to as \textit{optical pulling force} but originated from a different 
mechanism \cite{Novitzky1, Sukov, Ruffner, Novitzky2}.\\
In the present paper investigate NRP in case of CNLSE. The mechanism responsible for NRP presented in the present paper is different from the 
mechanism presented in our previous papers.\\
To the best of our knowledge it is possible to prepare an experiment which can test our predictions.

\textit{~~The model.~~}
We discuss a two component system described by coupled nonlinear Schr\"odinger equation (CNLSE) (symmetric interaction)
\begin{equation}\label{eq:model1}
i\partial_t\tilde\psi_i +\frac{1}{2}\partial_{xx}\tilde\psi_i+\left(\alpha_i^2\abs{\tilde\psi_i^2}+\gamma\abs{\tilde\psi_j^2}\right)\tilde\psi_i=0,
\end{equation} 
or after rescaling ($\alpha_i\tilde\psi_i=\psi_i$ and $\beta_i=\gamma/\alpha_i^2$)
\begin{equation}\label{eq:model}
i\partial_t\psi_i +\frac{1}{2}\partial_{xx}\psi_i+\left(\sgn(\alpha_i^2)\abs{\psi_i^2}+\beta_i\abs{\psi_j^2}\right)\psi_i=0,
\end{equation} 
where $i,j=1,2$ but $i\neq j$ and  $\alpha_i^2>0$. Following \cite{Goodman} we will use the second form. For special choice of 
$\beta_1=\beta_2=1$ or when one of the components is absent everywhere the system is integrable. 
There exists a particular solution, the soliton, which can be parametrized by $(\psi_1,\psi_2)=(e^{i\Omega_1 t}\Phi_1(x), e^{i\Omega_2 t}\Phi_2(x))$.
We will consider initially stationary solitons at $x=0$ described by even profiles $\Phi_i(-x)=\Phi_i(x)$ with amplitude $A_i=\Phi_i(0)$. 
The equation possesses scaling symmetry $\psi_i(x,t)\to\lambda\psi_i(\lambda^2t, \lambda x)$. We can use it to set $\gamma=1$. The solutions can be 
also multiplied by an arbitrary constant phase $\psi_i(x,t)\to e^{i\theta_i}\psi_i$.

\textit{~~Effective mass.~~}Momentum density in the model is defined as:
\begin{equation}
 \mathcal{P} = 
\frac{i}{2}\left(\Psi_x^\dagger\Psi-\Psi^\dagger\Psi_x\right),
\;\;\Psi=(\tilde\psi_1, \tilde\psi_2)^T
 \end{equation}
 or 
 \begin{equation}
   \mathcal{P} = \frac{i}{2}\sum_{k=1}^2\frac{1}{{\beta_k}}(\partial_x\psi_k^*\psi_k-\psi_k^*\partial_x\psi_k).
 \end{equation} 
The moving soliton can be written as
\begin{equation}
 \psi_i(x,t) = e^{i\Omega_it}\Phi_i(x-vt)e^{iv(x-t/2)}.
\end{equation} 
It's total momentum is equal to
\begin{equation}
 P = v\int_{\mathbb{R}}\left(\abs{\Phi_1^2}/\beta_1+\abs{\Phi_2^2}/\beta_2\right).
\end{equation} 
Another important property is the continuity equation $\partial_t\abs{\psi_i^2}/\beta_i+\partial_x\mathcal{P}_i=0$.

\textit{~~A wave.~~} The second 
type of solutions we discuss is the wave. 
In the absence of the soliton the wave with small amplitude in a single component can be 
written as $\psi_i = a_i\exp(i(\omega_i t+k_i x)$, where $k_i^2/2=-\omega_i$. 
Note that only negative frequencies describe propagating waves. 
The momentum density of the wave is $\mathcal{P}=ka^2/\beta$ which propagates with the velocity $v=-\omega/k$. 
If such a wave is escaping a box it carries away momentum equal to $P_t=-\omega a^2/\beta=k^2a^2/(2\beta)$. 
In the paper we consider a small amplitude $a_l\ll1$ wave incoming from $-\infty$ in the $l$-th sector with the following asymptotics
\begin{equation}\label{eq:bc1}
 \ppsi{i}{1}(x,t)\tendsto{x\to-\infty}a_le^{i\omega_it}\left(\delta_{li}e^{ik_ix}+R_ie^{-ik_ix}\right),
\end{equation} 
where $R_{li}$ is a reflection coefficient in the $i$-th channel and
\begin{equation}\label{eq:bc2}
 \ppsi{i}{1}(x,t)\tendsto{x\to\infty}a_lT_{li}e^{i\omega_i t+ik_{i}x}, 
\end{equation} 
with $T_{li}$ being the transition coefficient.
There is only one wave which brings in the momentum and the other waves carry away the momentum. 
The surplus which is left is the force exerted on the 
soliton
\begin{equation}
 P_t=F=a_l^2\sum_{i=1}^2\left(\delta_{li}+\abs{R_{li}^2}-\abs{T_{li}^2}\right)\abs{\omega_i}/\beta_i.
\end{equation} 
\textit{~~Perturbation series.~~} 
Assuming the wave has a small amplitude we can apply the perturbation theory with the wave amplitude as an expansion parameter.
Let us substitute
\begin{equation}
 \psi_i = e^{i\Omega_it}\sum_{n=0}^\infty a_l^n\psi_i^{(n)}\approx e^{i\Omega_i t}\left(\Phi_i+a_l\xi_i+\cdots\right),
\end{equation} 
where $\Phi_i$ is an $i$-th component of the stationary solution with frequency $\Omega_i$.
All the equations in $n$-th order will have the following form
\begin{equation}\label{eq:setODE0}
 \mathbf{L}\Psi^{(n)}=g^{(n)}\left(\Psi^{(0)},\ldots,\Psi^{(n-1)}\right),
 \end{equation} 
where $\Psi^{(n)} = \left(\psi^{(n)}_i,\,{\psi^{(n)}_i}^*,\,\psi^{(n)}_j,\,{\psi^{(n)}_j}^*\right)^T$
and
\begin{widetext}
 \begin{equation}\label{eq:setODE1}
  \mathbf{L} = \left(i\partial_t+\frac12\partial_{xx}\right)\mathbb{1}_4 -
 \left(\begin{array}{cc}
    \Omega_1&\\
    &\Omega_2
 \end{array}\right)\otimes\mathbb{1}_2 +\left(\begin{array}{cccc}
  2\Phi_1^2+\beta_1\Phi_2^2&	\Phi_1^2 &			\beta_1\Phi_1\Phi_2 &		\beta_1\Phi_1\Phi_2\\
  \Phi_1^2&			2\Phi_1^2+\beta_1\Phi_2^2 &	\beta_1\Phi_1\Phi_2 &		\beta_1\Phi_1\Phi_2\\
  \beta_2\Phi_1\Phi_2 &		\beta_2\Phi_1\Phi_2&		2\Phi_2^2+\beta_2\Phi_1^2 &	\Phi_2^2\\
  \beta_2\Phi_1\Phi_2 &		\beta_2\Phi_1\Phi_2&		\Phi_2^2&			2\Phi_2^2+\beta_2\Phi_1^2
  \end{array}\right).
 \end{equation} 
 \end{widetext}
Concerning a monochromatic wave in one of the components we can consistently decompose the first order solution into Fourier modes
 \begin{equation}
 \psi_i^{(1)}(x,t)=\xi_i(x,t)=e^{i\tilde\omega t}\xi^+_i(x)+e^{-i\tilde\omega t}\xi^-_i(x).
\end{equation} 
and rewrite the set of equations as 
\begin{equation}
 \left(\mathbf{K}+\mathbf{M}\right)\Xi=0,\qquad \Xi = \left[\xi_1^+, {\xi_1^-}^*, \xi_2^+, {\xi_2^-}^*\right]^T
\end{equation}
where
\begin{equation}\label{eq:kinetic}
 \mathbf{K} = \frac12\partial_{xx}\mathbb{1}_4 -  \diag\left(\Omega_1+\tilde\omega, \Omega_1-\tilde\omega, \Omega_2+\tilde\omega, 
\Omega_2-\tilde\omega\right).
\end{equation}
and $\mathbf{M}$ is the large matrix from (\ref{eq:setODE1}).
For a wave moving in the first component $\omega=\Omega_1-\tilde\omega<0$ and for the wave in the second component we will assume 
$\omega_2=\Omega_2-\tilde\omega<0$. The other two frequencies are positive and cannot propagate. 
We will solve the above system of equations with appropriate boundary conditions (\ref{eq:bc1}) and (\ref{eq:bc2}).
In general the above system has to be solved numerically but first we discuss two special cases 
of a single component soliton and a wave in one of the components.

~~\textit{Scalar soliton.}~~
Let us consider the scalar soliton in the first sector ($\Phi_1(x)=A_1\sech(A_1x)$, $\Phi_2(x)=0$) with the phase rotating with the frequency 
$\Omega_1=A_1^2/2$. 
The momentum of such soliton moving with velocity $v$ is equal to $P=2 A_1v/\beta_1$ therefore we can identify its mass as $m=2
A_1/\beta_1$.
First let us assume that the wave is present only in the first component ($\xi_2=0$) or in more general we seek solutions with $\psi_2\equiv 0$. 
The equation in the linear order can be reduced and written in a matrix form as $\mathbf{B}\left(\xi_1^+, {\xi^-_1}^*\right)^T=0$ with
\begin{equation}
 \mathbf{B} = \left(-\Omega+\frac{1}{2}\partial_{xx}+2\Phi^2\right)\mathbb{1}+
  \left(\begin{array}{cc}
  \tilde\omega& \Phi^2\\
   \Phi^2&-\tilde\omega
       \end{array}
\right).
\end{equation} 
Surprisingly, because of the integrability, the solution can be found quite easily in a closed form
\begin{equation}
 \ppsi{1}1(x,t)=e^{i\Omega_1 t}\left[ \frac{e^{i(\tilde\omega t+kx)}}{\cosh^2(x)} +e^{-i(\tilde\omega t+kx)}(k-i\tanh(x))^2 \right].
\end{equation} 
The first term, oscillating with the frequency $\Omega_1+\tilde\omega>0$ describes localized, nonpropagating correction. 
The second term with frequency $\omega_1=\Omega_1-\tilde\omega<0$ describes a moving wave. 
It is worth noting the wave moves in one and only in one direction and has no reflected part. 
We believe this is the feature of all integrable models. 
There is no dynamical interaction (no energy exchange)  between the soliton and the wave. 
The soliton is transparent to the radiation. One can only observe a phase shift. 
The same feature was also seen in sine-Gordon (sG) and (nonitegrable) $\phi^4$ model. 
sG kinks are transparent in all orders of the perturbation series. 
We believe this is also the case for CNLSE. 
Our numerical simulations of the full nonlinear equation (\ref{eq:model}) confirmed that the soliton does not feel any force exerted by the wave.

\begin{figure}
 \begin{center}
  \includegraphics[width=\columnwidth]{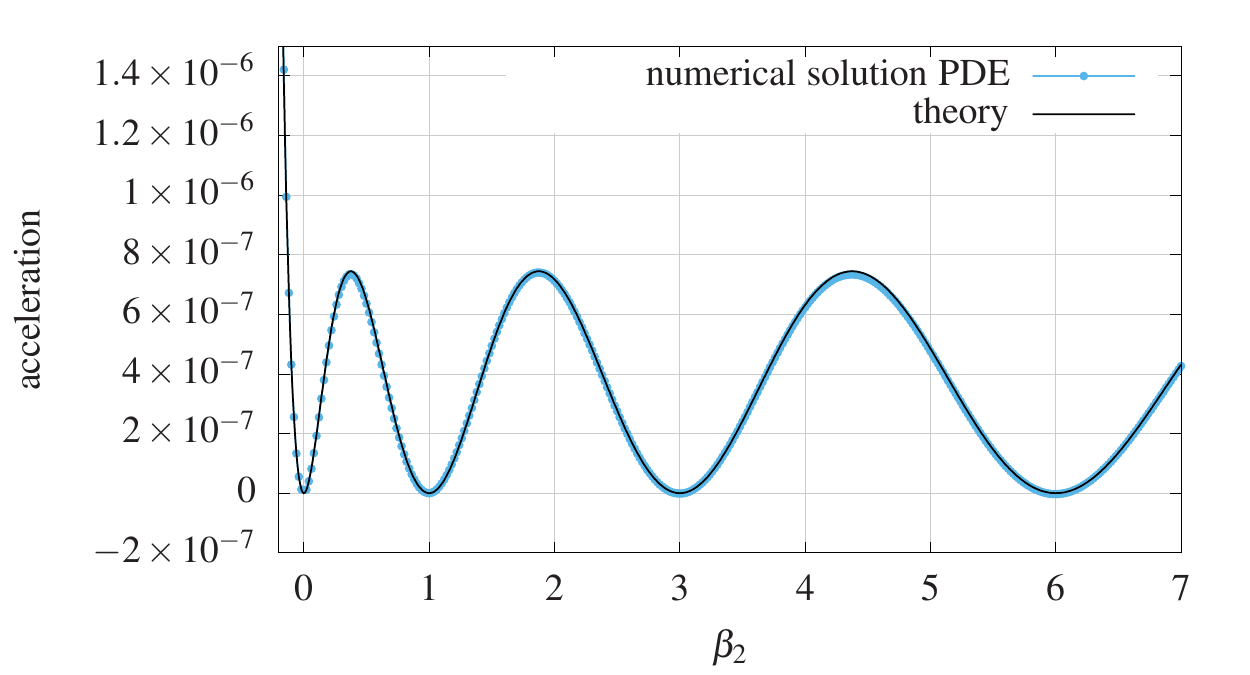}
  \caption{\small Measured (blue points) and calculated acceleration of the soliton for $A_1=1,\,A_2=0$ and $a_1=0,\,a_2=0.1$ and 
$\omega_2=-0.5$.}\label{fig:accel1}  
 \end{center}
 \end{figure}

\textit{~~Scalar soliton and a wave in the second component.}~~Let us consider again the scalar soliton in the first component but with the 
incident wave propagating in the second sector. 
Still  $\Phi_2(x)=0$ but $\ppsi11=0$.
Equations for  positive $\ppsi{2,+1}{1}$ and negative frequency $\ppsi{2,-1}{1}$ solutions decouple. 
We can set $\ppsi{2,+1}{1}=0$ as positive frequencies cannot propagate.
Equation for $\ppsi{2}{1}=e^{-i\omega t}\xi_2^-(x)$ now can be reduced to a single ordinary differential equation
\begin{equation}\label{eq:PTpotential}
  -\omega\xi_2^-+\frac{1}{2}\partial_{xx}\xi_2^-+\beta_2{\sech}^2(x)\,\xi_2^-=0.
\end{equation} 
Soliton from the first component is a source of a potential over which the wave in the second component scatters. 
The potential is a well known P\"oschl-Teller potential and has known solutions which in terms of of Legendre functions which can be written as
\begin{equation}
 \ppsi{1}{1,1}=e^{-i\omega t}P^{ik}_\lambda(\tanh x),\;\lambda=\pm\frac{1}{2}\sqrt{8\beta_2+1}-\frac{1}{2}
\end{equation} 
Note that for $\beta_2=\lambda(\lambda+1)/2$ for integer $\lambda$ the potential is reflectionless (F=0).\\
The asymptotics of the Legendre functions as $|x|\to\infty$ are well known in terms of gamma function. 
Imposing the boundary condition of incoming wave from $+\infty$ with amplitude equal to $1$ ($\xi_2^-(x)\tendsto{x\to-\infty}e^{ikx}+Re^{-ikx}$) we 
read the reflection coefficient which is needed to calculate the momentum balance and hence the force acting on the soliton 
$F=-k^2|R^2|a_2^2/\beta_2$ from which we calculate the acceleration
\begin{equation}
\ddot x = 
\frac{a_2^2k^2\sin^2(\lambda\pi)}{\sin^2(\lambda\pi)\cos^2\left(\frac{k\pi}{A}\right)+\cos^2(\lambda\pi)\sin^2\left(\frac{k\pi}{A}\right)}\frac{
\beta_1 }{\beta_2}.
\end{equation} 
Note that the direction of the acceleration is determined by the sign of the quotient $\beta_1/\beta_2$.
If both has the same sign the soliton experiences Positive Radiation Pressure (PRP) and if $\beta$s have opposite signs the soliton undergoes 
Negative Radiation Pressure (NRP).
This works perfectly, see Fig.\ref{fig:accel1}. 
By changing the sign of $\beta_1$ we observed numerically that the radiation pressure changes the sign as well. 
However, for $\beta$s with the different signs the equation (\ref{eq:model}) cannot be transformed into (\ref{eq:model1}) and it is possible that 
this 
NRP result has no physical meaning in this context.

\begin{figure}
 \flushleft{\tiny \ \ \ \ a)\hspace*{0.48\linewidth}b)}\\\vspace*{-0.35em}
 \includegraphics[width=0.49\linewidth,angle=0]{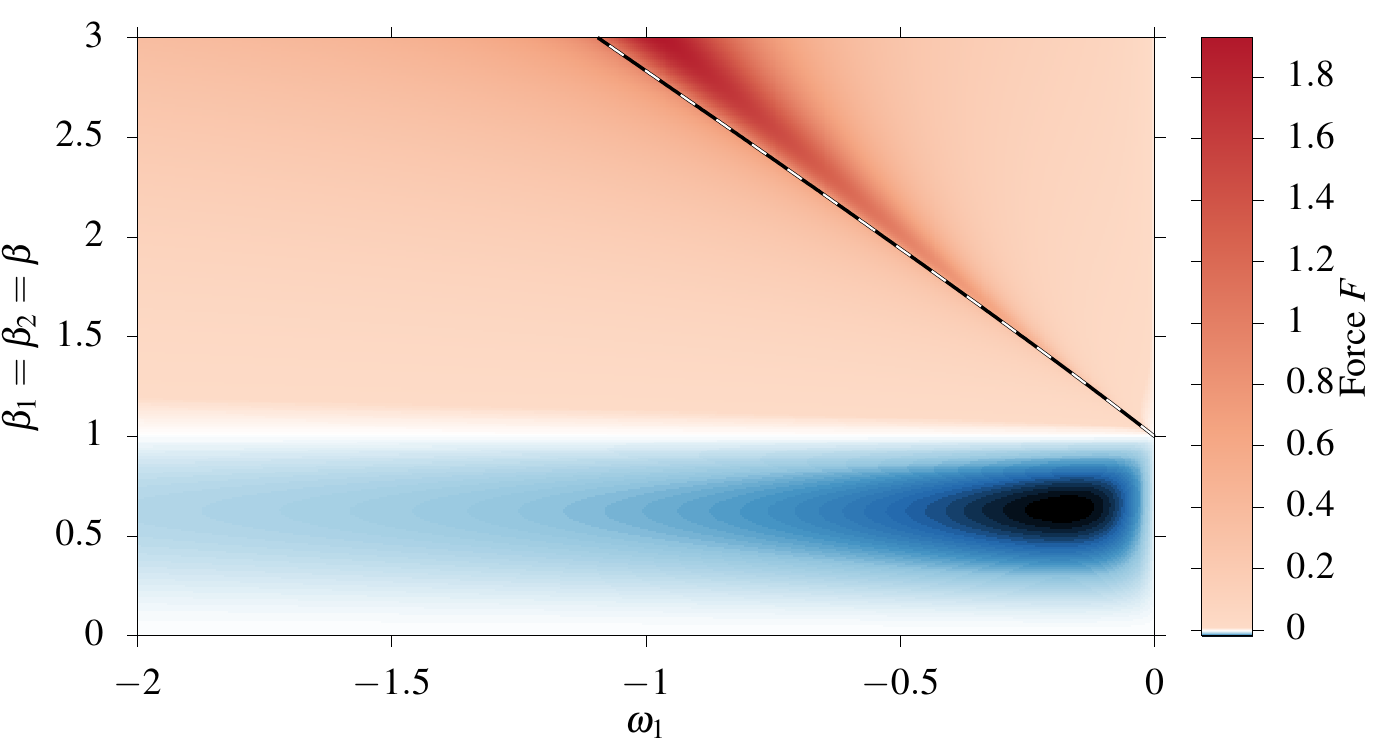} \includegraphics[width=0.49\linewidth,angle=0]{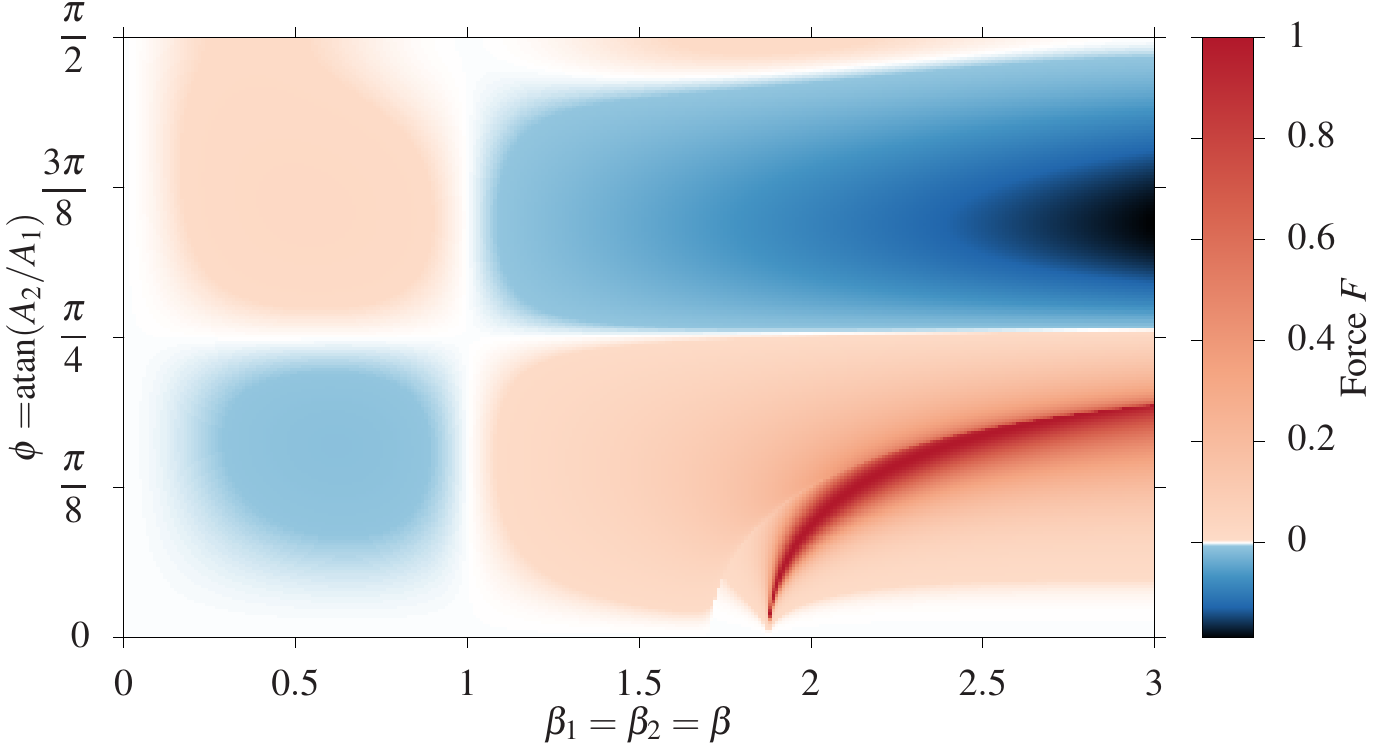}
 \vspace*{-3em}
 \flushleft{\tiny  \ \ \ \ c)\hspace*{0.48\linewidth}d)}\\\vspace*{-0.5em}
 \includegraphics[width=0.49\linewidth,angle=0]{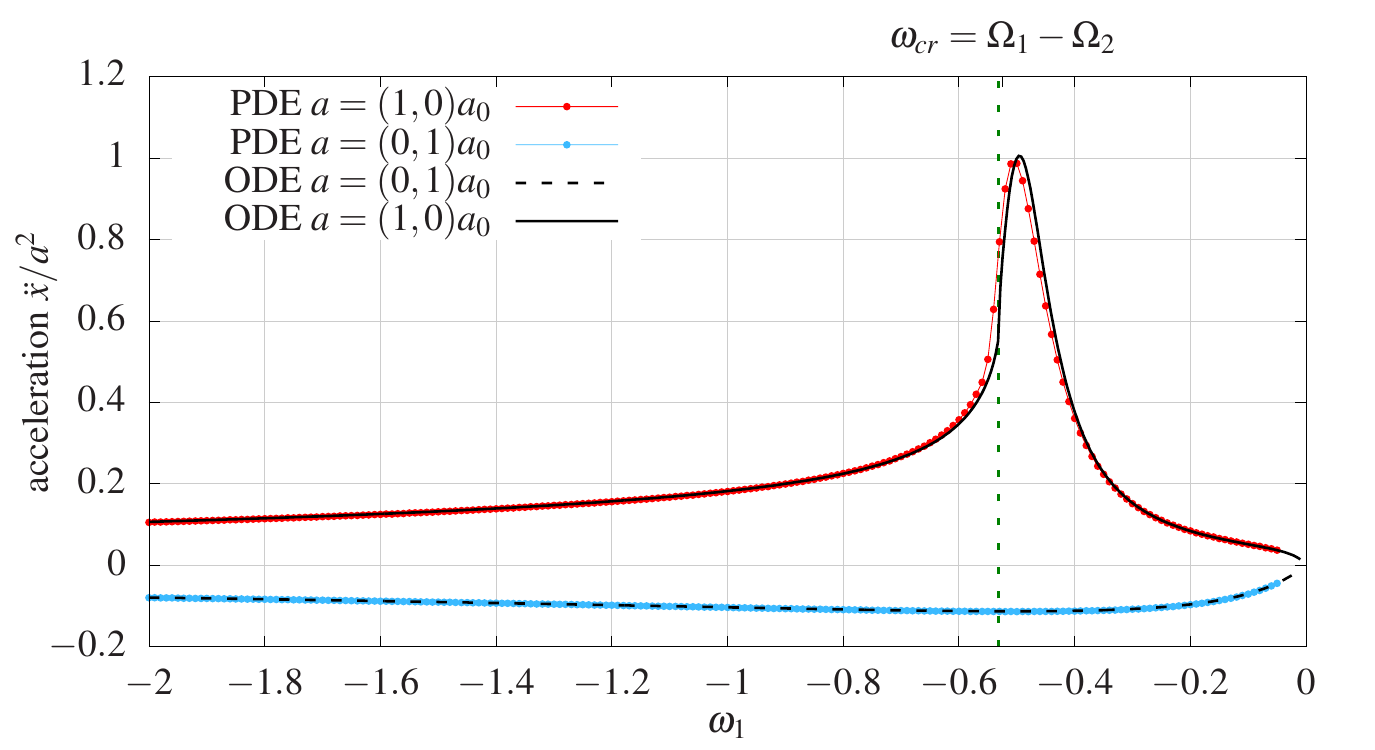}\includegraphics[width=0.49\linewidth,angle=0]{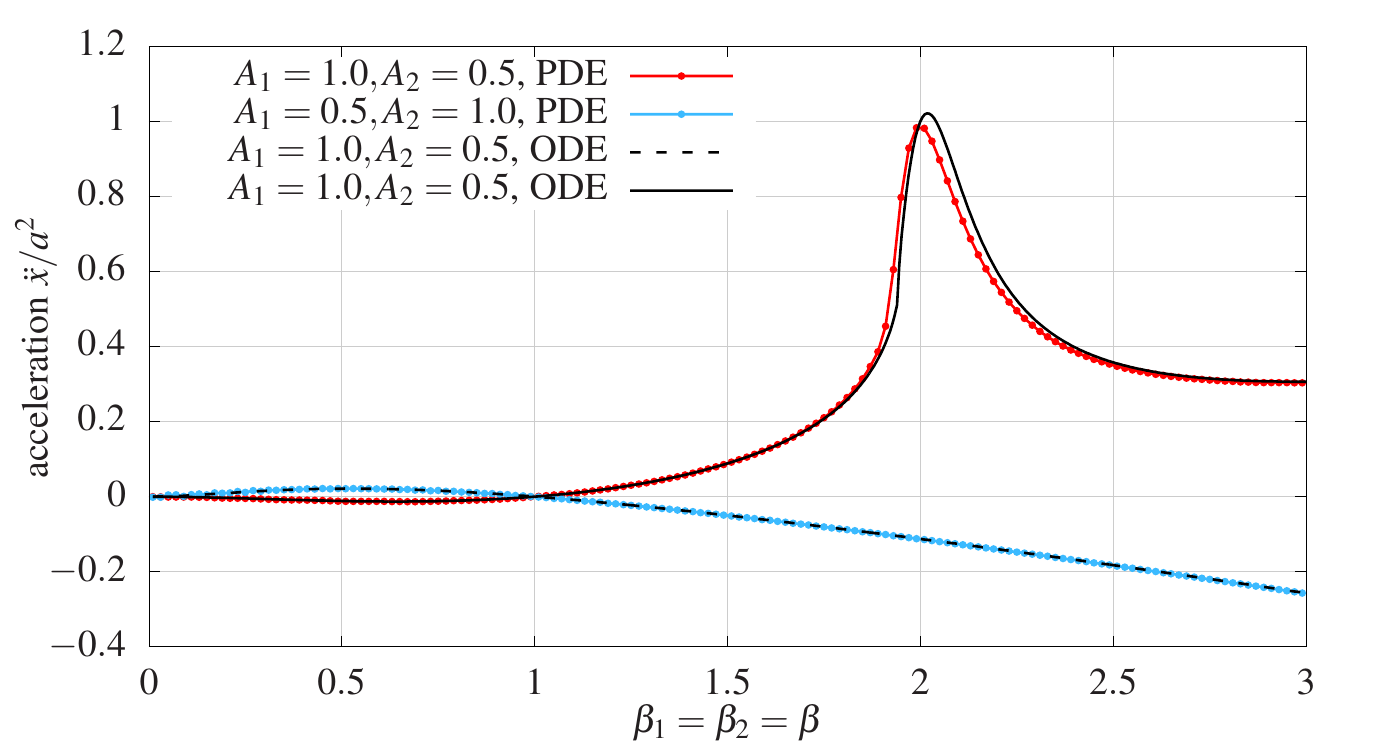}
 \caption{\small a) Force acting on soliton calculated from the solutions of (\ref{eq:setODE1}) for $A_2=0.5$ and $A_1=1$ as a function of $\beta_1$ 
and $\omega_1$. NRP visible as blue for $\beta_1<1$. Critical frequency is shown with dashed line.\\
b) Force acting on soliton calculated from the solutions of (\ref{eq:setODE1}) for $\omega=-0.5$ as a function of $\beta_1$ and 
$A_1=\cos\phi,\,A_2=\sin\phi$. NRP visible as blue for $\beta_1<1$ and $A_2<A_1$ and for $\beta_1>1$ and $A_2>A_1$.\\
c)  Acceleration calculated from linearized problem (ODE) compared with the full PDE solution for $\beta_1=\beta_2=2, \,A_1=1,\,A_2=0.5$ and 
$a_0=0.01$ as a function of wave frequency $\omega$. Solid line shows the acceleration for the wave in the first sector, dashed in the second.\\
d) Acceleration calculated from linearized problem (ODE) compared with the full PDE solution for $\omega=-0.5$ and $a_1=0.01$ as a function of 
wave frequency $\beta_1=\beta_2$. }\label{fig:calculations}
\end{figure}

~~\textit{Vector soliton}~~Let us now consider a monochromatic wave in the first sector hitting the two component vector soliton. 
$A_2=0$ corresponds to the previous case and when $A_2$ tends to infinity it formally reproduces the first, integrable case. 
In a special case, when $\beta_1=\beta_2=\beta$ and the solitons have the same amplitude we can also find their profiles and frequencies 
analytically $\Phi_1=\Phi_2=A\sech(\sqrt{1+\beta}Ax),\Omega=\sqrt{1+\beta}A^2/2$. The total momentum in this case equals 
$P=\frac{4Av}{\beta\sqrt{1+\beta}}$ so the mass of the 
vector soliton is equal to $4A/(\beta\sqrt{1+\beta})$.
In the case when the second component is much smaller $A_2\ll A_1=1$ it can be treated as a linear perturbation to the large soliton and is ofter 
referred to as a \textit{shadow soliton}. 
In such a case the amplitude of the second soliton is arbitrary but small, and the frequency of the companion soliton does not depend on the amplitude (in the first approximation)
$\Phi_1=\sech(x),\Omega_1=1/2$, $\Phi_2=A_2\sech^\lambda(x),\; \Omega_2=\lambda^2/2$. Formally $\Phi_2$ satisfies equation 
(\ref{eq:PTpotential}) but with positive 
eigenfrequency and vanishing boundary conditions as $\abs{x}\to\infty$.  There can be other higher bound states of the potential (depending on 
$\beta_2$). When $A_2>0$ they couple with the scattering modes of the field $\psi_1$  and become resonance modes.\\
In general the profiles and the frequencies of the soliton counterparts are to be determined numerically by solving the system of two nonlinear ODEs 
with appropriate boundary conditions, but the frequencies 
for $\beta_1=\beta_2=\beta$ the frequencies change monotonically from $(\Omega_1,\Omega_2)=(1/2, \lambda^2/2)$ as $A_2\to 0$ to $(1/\sqrt{1+\beta}, 
1/\sqrt{1+\beta})$ for $A_2=1$. \\
Let us first consider a wave $\xi_1^{-}$  propagating in the first sector with the frequency $\omega_1=\Omega_1-\tilde\omega$.
It  is accompanied by the nonpropagating localized perturbation  $\xi_1^{+}$ oscillating with $\Omega+\tilde\omega=2\Omega_1-\omega_1$. 
In the second sector the nonpropagating perturbation $\xi_2^{+}$ has the frequency $\Omega_2+\tilde\omega=\Omega_2+\Omega_1-\omega_1$. 
The second perturbation  $\xi_2^{-}$ oscillates with $\omega_2=\Omega_2+\tilde\omega=\Omega_2-\Omega_1+\omega_1$. 
The most important dynamical properties of this scattering problem depend on this frequency.
For simplicity let us assume that the reflection is negligible. From the numerical solution we can see that the reflection is much smaller that the 
transition for wide range of parameters. We can now discuss only the simplified scattering problem from $\xi^{+}_1$ to $\xi^+_1$ and $\xi^+_2$. 
If $\abs{\omega_2}>\abs{\omega_1}$ (which is true when $\Omega_2<\Omega_1$) the wave excited in the second component moves faster than initial wave 
and hence carries more momentum. This surplus has to be balance by additional force acting on the soliton. In this case the force pushes the soliton 
backwards which can be interpreted as the \textit{Negative Radiation Pressure}. 
Let us stress again that this is true only when the reflection is sufficiently small. However this feature was confirmed by both solving the 
linearized system of ODEs (\ref{eq:setODE1}) and the full system of nonlinear PDEs (\ref{eq:model}).
When $\abs{\omega_2}<\abs{\omega_1}$ meaning $\Omega_2>\Omega_1$ the scattered wave in the second component carries less momentum, and the soliton is 
pushed by the wave revealing \textit{Positive Radiation Pressure}. 
Moreover there is a  critical value of $\omega_1=\omega_\text{cr}$ when $\omega_2=0$. 
Above this frequency, the second companion cannot propagate, and only one channel of scattering remains open. 
At $\omega_1=\omega_\text{cr}$ the kinetic term in the linearized system has a secular term meaning that the Fourier decomposition doesn't work in 
this case. 
Near that critical value the force acting on the soliton becomes large, showing resonant behavior. 
These predictions were verified by soling the linearized equation (\ref{eq:setODE0}) for some set of parameters. 
For constant amplitudes $A_1=1$ and $A_2$ we varied $\beta_1=\beta_2=\beta$ and frequency $\omega_1$. 
For $\beta<1$ ($\Omega_2<\Omega_1$) we generally obtained small but negative force and for $\beta>1$ ($\Omega_2>\Omega_1$) the force was always 
positive (Fig.~\ref{fig:calculations}a). %(Fig.~\ref{fig.betaOmegaScan}).
For constant frequency $\omega_1=-0.5$ we also varied $\beta_1=\beta_2=\beta$ and the polarization of the soliton but keeping $A_1^2+A_2^2=1$ 
constant.
For $A_1<A_2$ and $\beta<1$ or $A_1>A_2$ but with $\beta>1$ the frequency of the second companion was larger $\Omega_2>\Omega_1$ and the force was 
positive. For $A_1>A_2$ and $\beta<1$ or $A_1<A_2$ but with $\beta>1$ ($\Omega_2<\Omega_1$) and when the reflection was small enough negative force 
was obtained (Fig.~\ref{fig:calculations}b). %(Fig.~\ref{fig.betaA2}). 
Especially on this figure but to some extent also on the Fig.~\ref{fig:calculations}a %~\ref{fig.betaOmegaScan} 
one can see two lines along which the force changes very 
rapidly. 
One of them is easily to identify as the critical frequency.
The second line corresponds to the first resonance or quasi-normal mode which we believe should be present for $\beta>1$ and small 
values of $A_2$.
We have also performed a full numerical simulation of the PDE (\ref{eq:model}). 
Example comparison between the acceleration measured from the full nonlinear numerics and the linearized analysis  
is shown on Fig.~\ref{fig:calculations}c. %Fig.~\ref{fig.PDEODEComparizon}. 
We have found that the agreement is sufficient only for large frequencies. 
Near the critical frequency the linear approximation breaks.

\begin{figure}
\centering
\includegraphics[width=\linewidth,angle=0]{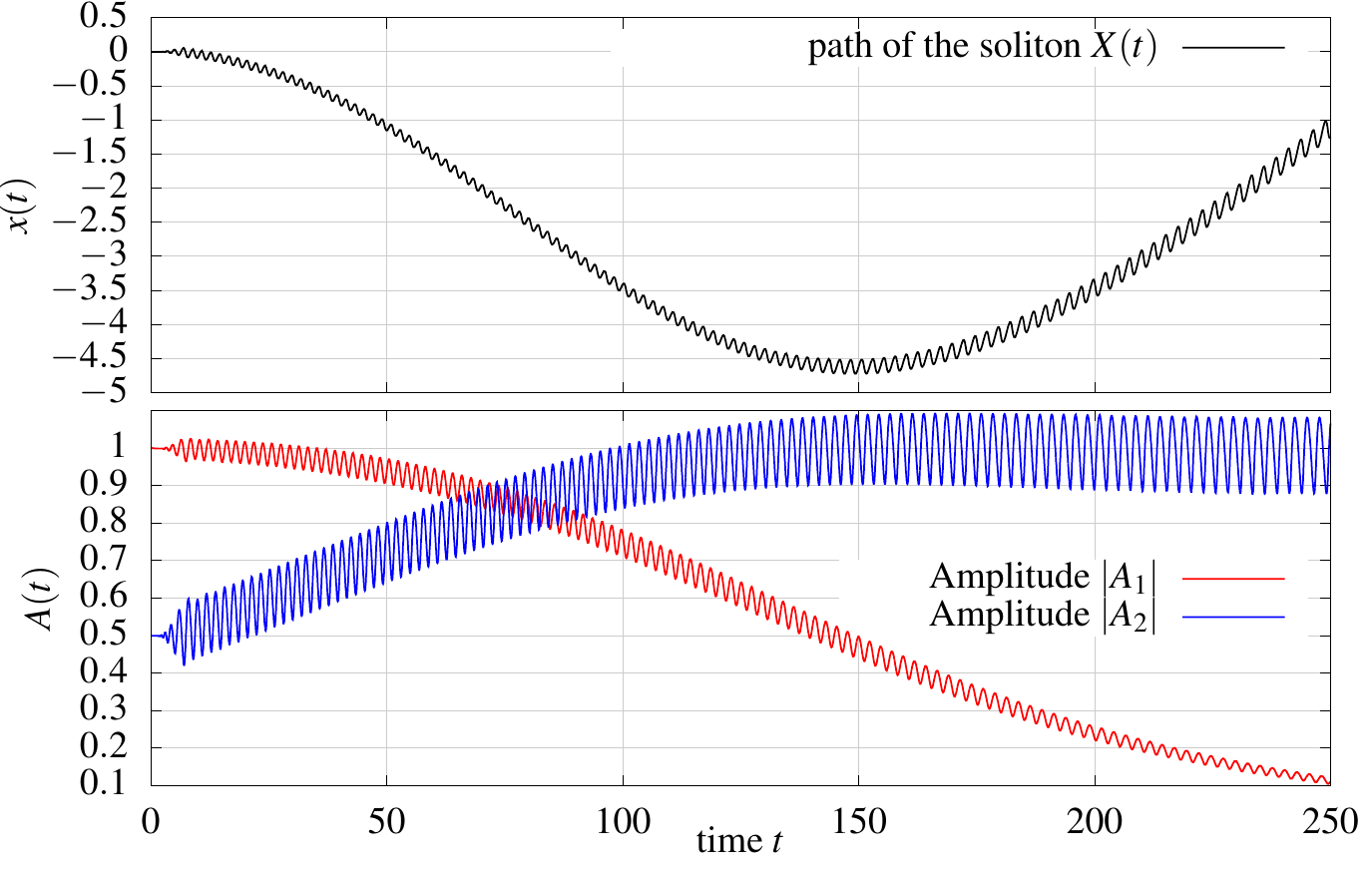}
\caption{\small Path of the soliton and the evolution of its amplitude under the influence of a large wave $a_2=0.1$ for $\omega=2$, 
$\beta=2$, $A_1=1$, $A_2=0.5$. Initially the soliton is pulled by the wave. But in time the amplitude change and when $A_1$ becomes smaller than 
$A_2$ the acceleration changes its direction.}\label{fig.paths}
\end{figure}

~~\textit{Long time behavior.}~~ Note that the solution to the linear equation satisfying the boundary conditions 
(\ref{eq:bc1}) and (\ref{eq:bc2}) violates the continuity equation. 
It is especially obvious in the sector without the incoming wave. 
The continuity equation states that $|R_{12}^2|+|T_{12}^2|=0$. 
In fact this is not a contradiction because the continuity equation can be satisfied using higher order terms. 
However one important conclusion can be drown here already. 
Physically the wave created by the scattering carries away particles and energy from the soliton (forced emission). 
In order to create a wave in the sector without incoming wave the amplitude of the soliton must decrease. 
This is also what can be observed in the full numerics (Fig.~\ref{fig.paths}). 
Probably higher order analysis would reveal this feature from the perturbation series.
It is worth noting that similar secular terms to that near the critical frequency are present also in higher orders of the perturbation series for 
any frequency. 
The source term  $g^{(2)}$ contains terms like $\psi_1^{(1)}{\psi_1^{(1)}}^*$ which oscillate with the frequency of the soliton. 
This term does not depend on the frequency $\tilde\omega$ because the complex conjugation kills that dependence. 
Therefore it cannot be canceled by any cancellation of the resonant terms procedure. % such as presented by \cite{LandauLifszits}. 
The meaning of this feature is that the Fourier decomposition may work only for a very short time, before the resonant terms become important.
We believe that these resonant terms are responsible for the changes of amplitude of the soliton.

~~\textit{Conclusions.}~~In the paper we have shown that solitons in CNLSE show very rich and sometimes unexpected interaction with small waves.
The most important conclusion is that the solitons can experience  both positive and negative radiation pressure depending on the  sector in which 
the incident wave travels. 
Scattering from faster rotating counterpart can result in NRP. 
Scattering from slower to faster always results PRP. 
We have also found that the wave can take away particles from the soliton leading to the decrease of its amplitude very similar to the forced 
emission known from atomic physics. This effect is very different from interactions between topological solitons and waves studied in our earlier 
papers.\\ 
Although CNLSE is nonrelativistic equation we expect to see similar phenomenon also in case of the multicomponent Q-Balls which can have importance 
in cosmological context. Scattering on Q-balls can also happen between modes with different frequencies and momenta.\\

~~\textit{Acknowledgements.}~~The author wants to express his gratitude for Krzysztof Sacha for many hours of inspiring and enlightening discussions.

~~\textit{Supplementary material.}~~The paper has additional material in form of four short animations depicting the problems discussed.
\newcommand\movitem[1]{\item\hspace*{-20pt}\texttt{#1}\,:\,}\medskip
\begin{enumerate}[label=]
 \movitem{beta32\_first.mov} Vector soliton $\beta=3/2$ with the wave in the first sector experiencing PRP and decay of the second companion.
 \movitem{beta32\_second.mov} Vector soliton $\beta=3/2$ with the wave in the second sector experiencing NRP and pumping of the second companion.
 \movitem{betas1.mov} Soliton in the first sector and the wave in the second. $\beta$s with the same signs (PRP).
 \movitem{betas2.mov} Soliton in the first sector and the wave in the second. $\beta$s with different signs (NRP).
\end{enumerate}
\begin{small}

\end{small}
\end{document}